\documentstyle[12pt,bezier]{article}

\newcommand{\dpar}[2]{\frac{\partial #1}{\partial #2}}
\newcommand{\vpar}[2]{\frac{\delta #1}{\delta #2}}
\newcommand{\ddpar}[2]{\frac{\partial^2 #1}{\partial #2^2}}

\newcommand{\bm}[1]{\mbox{\boldmath $#1$}}
\def\ltap{\raisebox{-.55ex}{\rlap{$\sim$}} \raisebox{.4ex}{$<$}}
\def\gtap{\raisebox{-.55ex}{\rlap{$\sim$}} \raisebox{.4ex}{$>$}}

\def\e{\mbox{e}}

\def\arccosh{\mbox{arccosh}}
\def\half{{1 \over 2}}

\begin{document}
\title{Instanton--Like Transitions  at High Energies
       in (1+1) Dimensional Scalar Models}
\author{
   D.~T.~Son\\
  {\small \em Institute for Nuclear Research of the Russian Academy of
  Sciences,}\\
  {\small \em 60th October Anniversary prospect 7a, Moscow 117312}\\
    and\\
   V.~A.~Rubakov\\
    {\small \em Institute for Nuclear Research of the Russian Academy of
  Sciences,}\\
  {\small \em 60th October Anniversary prospect 7a, Moscow 117312}\\
  {\small and}\\
  {\small \em Department of Physics and Astronomy, Rutgers
  University,}\\
  {\small \em Piscataway, New Jersey, 08855}\\
  }
\date{October 1993}
\maketitle
\begin{abstract}
Instanton-like transitions (``shadow processes'') are considered in
 (1+1) dimensional models with one scalar field whose potential
is a  quadratic well with a cliff. The corresponding classical
boundary value problem is solved, and
the semiclassical transition probabilities are found in a rather wide
 range of energies and the
numbers of initial particles.

\end{abstract}

\newpage

\section{Introduction}

Much effort is being made to understand the high energy behavior of
the amplitudes of instanton--like processes. Being suppressed at zero energy
by the factor $\e^{-S_0}$, where $S_0$ is the action of the instanton, these
amplitudes exhibit the exponential growth with energy, and, in the leading
order of the perturbation theory
around the instanton, cross the unitarity bound, typically at the sphaleron
mass scale \cite{Ringwald}. This observation has lead to an exciting
speculation that the electroweak baryon--number violating processes can
become observable in the TeV energy region \cite{Ringwald,MVV}
 (for reviews see refs.\cite{Mrev,Trev}).

There are strong arguments implying
that the behavior of the instanton--like cross
section is exponential,
\begin{equation}
  \sigma=\exp\left[-{1\over g^2}F\left(E\over E_0\right)\right],
  \label{FE}
\end{equation}
where the energy scale $E_0$ is of the order of the sphaleron mass in the
electroweak theory and many other models; the function $F$ is the
central object of current studies. Until now, only low energy asymptotics of
$F(E/E_0)$ is understood, and a few terms of the corresponding series have
been explicitly calculated within the perturbation theory around the
instanton \cite{Zakharov,KRT,Porrati,KhozeRingwald,MuellerDiakonovPetrov}.
The peculiar form of the cross section suggests that even at energies
comparable to the sphaleron mass, where the function $F$ cannot be found by
the perturbation theory around the instanton, there might exist a
semiclassical--type procedure for calculating the amplitudes. The key problem
for such a procedure is the non--semiclassical
nature of the two initial hard particles \cite{Mueller2}.

Recently, several approaches that may allow to overcome this problem have been
proposed \cite{RT,Tinyakov,Khlebnikov,DPN,VoloshinN}.
One of them
\cite{RT,Tinyakov} is as follows.
  Instead of considering two--particle scattering, one examines the
maximum probability of transitions among all initial states with given energy
$E$ and number of initial particles $n=\nu/g^2$, where $\nu$ is fixed in
the limit $g^2\to 0$. In this regime, the total
probability
indeed has the exponential form (\ref{FE}), where the function $F$
now depends on $\nu$. At finite $\nu$, there exists an entirely semiclassical
procedure for calculating the function $F(E/E_0,\nu)$ which,
in its final version,  is reduced
to a well--defined classical boundary
value problem \cite{RST}. Though not a priori justified, a rather natural
assumption is that the limit $\nu\to 0$ of the function $F(E/E_0,\nu)$ is
smooth
and its limiting value $F(E/E_0,\nu\to 0)$ gives at least the upper bound
for, and may even coincide with the two--particle function $F(E/E_0)$. The
latter conjecture has found substantial support from high--order
perturbative calculations around the instanton \cite{Mueller3}.

The prescription for the evaluation of the function $F(E/E_0,\nu)$ is as
follows (see ref.\cite{RST} for details). One searches for a solution to the
field equations,
\begin{equation}
  \vpar{S}{\phi}=0,
  \label{bvp1}
\end{equation}
on the contour  ABCD in the complex time plane which is shown in fig.1,
with the following boundary conditions,
\[
  \phi(\bm{k})={1\over\sqrt{2\omega_{\bf k}}}
               (f_{\bf k}\e^{-i\omega_kt'}
               +\e^{\theta}f^*_{-{\bf k}}\e^{i\omega_kt'})~~
  \mbox{when}~t'\to -\infty,
\]
\begin{equation}
  \phi(\bm{k})={1\over\sqrt{2\omega_{\bf k}}}
               (b_{\bf k}\e^{-i\omega_kt}+b^*_{-{\bf k}}\e^{i\omega_kt})~~
  \mbox{when}~t\to +\infty,
  \label{bvp2}
\end{equation}
where $T$ and $\theta$ are two free parameters of the field configuration
which are related to the energy and the number of initial particles,
$t'=t-iT/2$
is the real parameter on the part $AB$ of the contour, and $f_{\bf
k}$ and $b_{\bf k}$ are arbitrary functions of the spatial momentum $\bm{k}$.
In other words, the boundary conditions (\ref{bvp2}) require that the ratio
between the negative-- and positive--frequency parts of the asymptotics of
the field $\phi$ is equal to $\e^{-\theta}$ in the initial state, and 1 in the
final state. The solution to this boundary value problem is
complex on the contour of fig.1 (except for $\theta=0$), but is real on the
part $CD$. It describes the most probable ``microcanonical'' process among
all initial states with energy
\[
  E=\int d\bm{k}\omega_{\bf k}\e^{\theta}f^*_{\bf k}f_{\bf k}
   =\int d\bm{k}\omega_{\bf k}b^*_{\bf k}b_{\bf k}
\]
and number of particles
\[
  n=\int d\bm{k}\e^{\theta}f^*_{\bf k}f_{\bf k}.
\]
The classical action of the field evaluated along the contour of fig.1,
which we will denote by $iS$ (so that $S$ is real and positive for
purely Euclidean fields, such as an instanton), satisfies
the following relations,
\begin{equation}
  2\dpar{S}{T}=E,~~~ 2\dpar{S}{\theta}=n,
  \label{derivS}
\end{equation}
and the function $F$ is merely the Legendre transform of the double
action,
\begin{equation}
  {1\over g^2}F(E,n)=2S-ET-n\theta.
  \label{FS}
\end{equation}
Obviously, it satisfies the following relations,
\begin{equation}
  \dpar{(F/g^2)}{E}=-T,~~~ \dpar{(F/g^2)}{n}=-\theta.
  \label{derivF}
\end{equation}

It is likely that the above boundary value problem,
because of its complexity, can
be solved only by numerical methods, except for the region of low energies,
$E\ll E_0$, where $F$ can be found by the perturbation theory
around the instanton.  However, for
the qualitative understanding of the structure of the solution as well as the
behavior of the function $F(E/E_0,\nu)$, it is desirable to invent a simple
model where the solution and the function $F$ can be found analytically,
 at least in some non-trivial range of $E$ and $n$, by
techniques different from the usual perturbative expansion around a single
instanton.

In this paper we present such a model. We consider a (1+1)--dimensional
theory of one scalar field, whose  potential has a local minimum at $\phi=0$,
but is unbounded from below. The decay of the false vacuum $\phi=0$ is
described by a bounce configuration \cite{Coleman}, which shares many
properties of instantons in gauge theories. Making use of this bounce
solution, one can consider a ``shadow process'' \cite{Hsu,Voloshin},
 i.e. scattering of particles
through the formation of a bubble. The shadow process is a direct analog of
instanton--induced scattering and it can be treated by the same methods as
those used in gauge theories. Though not precisely defined at high energies,
the ``shadow process'' is a good choice for testing the semiclassical
technique proposed in ref.\cite{RST}.

In the model that we will study in this paper, there exists an additional
large parameter besides the inverse coupling constant. When this
parameter tends to infinity, the potential has the form of a quadratic
well with a cliff.  In this limit, the boundary value problem is
solvable in a rather wide range of $E/E_{sph}$ and $\nu$, and the solution is
a non--trivial generalization of the dilute instanton gas. The specific
models to be considered in this paper are described by the exponential
interaction, $\exp(\lambda\phi)$, and power--like one, $\phi^N$, where
$\lambda$ and $N$ are large.

For the two models, the results are qualitatively the same. The periodic
instanton, which is the solution to the boundary value problem at
$\theta=0$, can be found in the whole range of energies where it exists,
i.e., $0<E<E_{sph}$. For $\nu$ of order 1 the function $F$ can be found for
all energies smaller than some critical value $E_{crit}(\nu)$ where the
exponential suppression is reduced by a factor of order
 $\lambda^{-1}\ll 1$ (i.e., at the critical energy $F\sim S_0/\lambda$);
the smaller the number of initial particles, the larger the critical energy.
For even smaller $\nu$, namely, $\nu\sim\lambda^{-1}$, the approximations
made in this paper fail at relatively low energies, where the exponential
suppression is reduced by a factor of order one only. The limit $\nu\to 0$ is
smooth and in the exponential model the function $F(E/E_0)$ coincides with
the naively extrapolated one--instanton result, that shows that all higher
terms of the perturbative expansion around the instanton
 vanish. In the power model, the result does
not coincide with the one--instanton formula and corresponds to summing up
the whole perturbation series, where each term is calculated to the leading
order in $1/N$.

The fact that the function $F(E/E_0,\nu)$ can be calculated beyond the
perturbation theory is a remarkable feature of our model. Unfortunately, our
results for the most interesting case $\nu\to 0$ are rather limited (we can
only trust the reduction of the exponent by a factor $2/3$ and $\e^{-1/2}$ in
the exponential model and  power model, respectively), so they do not tell
much on whether the two--particle instanton--like scattering is strong or
not at high energies.

One question that can be addressed quantitatively in these models is whether
multi--instanton processes become important at those energies when
one--instanton amplitudes are still exponentially small. This conjecture
\cite{Zakharov1,M&S},
 that has lead to a hypothesis of premature
unitarization, is discussed in this paper within the exponential model. For
processes with relatively large number of initial particles, which we can
safely treat until the exponential suppression is strongly reduced (i.e.
until $E=E_{crit}(\nu)$), we find that multi--instanton contributions are
more suppressed than single--instanton ones at all energies up to $E_{crit}$.
We think this is a strong argument against the premature unitarization
hypothesis.

The paper is organized as  follows. Sects. 2 -- 5 are devoted to
the exponential model.
 In Sect.2 we describe the
exponential model and find the analogs of the sphaleron and the instanton.
Sect.3 is the central part of the paper and contains the construction of
what we call the ``improved dilute instanton gas approximation''. The
solution to the boundary value problem is found, and the function $F$ is
calculated for the two cases $\nu\sim 1$ and $\nu\sim 1/\lambda$. The
periodic instanton solution and multi--instanton contributions are also
discussed. In Sect.4 we try to go beyond the improved dilute instanton
gas and
describe the periodic instanton up to $E_{sph}$; unfortunately, we are
 unable to
obtain essentially new information on  the processes with small number of
initial particles at high energies beyond the improved dilute gas.
 Sect.5 is devoted to a related
problem, namely,  induced vacuum decay, which can be also considered as a
classical boundary value problem. The exponential suppression of
the vacuum decay is not reduced when the number of initial particles is
small, up to the energies where our approximation breaks down. The model with
$\phi^N$ interaction with large $N$, where our approach requires some
modifications, is considered in Sect.6. Sect.7 contains concluding remarks.

\section{The model}

\subsection{Shadow process}

The model to be considered in this paper, except for Sect.6, is a (1+1)
dimensional theory containing one scalar field $\phi$, with the lagrangian
\[
  L = {1\over 2}(\partial_{\mu}\phi)^2 - V(\phi).
\]
The potential $V(\phi)$ has a mass term and an exponential interaction term,
\begin{equation}
  V(\phi) =    {m^2\over 2}\phi^2 -
      {m^2v^2 \over 2}\exp\left[2\lambda\left({\phi\over v}-1\right)\right].
  \label{potential}
\end{equation}
Note that the interaction term has negative sign, so the potential is
unbounded from below at large positive $\phi$. The theory contains two
massless parameters, $v$ and $\lambda$. The parameter $1/v$ is the coupling
constant that governs the perturbation theory. The probabilities of
semiclassical processes we are interested in are exponential in $v$: in
general, expressions like eq.(\ref{FE}) will have the form
\[
  \sigma=\exp\left[-F\left(v^2;{E\over E_0},\nu;\lambda\right)\right]
        =\exp\left[-v^2f\left({E\over E_0},\nu;\lambda\right)\right],
\]
(i.e., $\ln \sigma\propto -v^2$ at fixed $E/E_0$ and $\nu$),
where $E_0\propto v^2$ and the number of initial particles is $n=v^2\nu$. In
this paper we  concentrate on the exponent $F$ and make no attempts to
estimate the pre--exponential factors. For the semiclassical expansion to be
consistent, we take $\lambda\ll v$. On the other hand, $\lambda$ itself
needs not be small, and we assume throughout this paper  that $\lambda\gg 1$.
So, we take
\[
  v\gg\lambda\gg 1.
\]

The behavior of the potential is shown in fig.2. Due to the
large value of the parameter $\lambda$, the potential has the form of a
quadratic well with a cliff. At $\phi=v$, the mass term and the
interaction term in the potential cancel each other, so $V(v)=0$. When $\phi$
decreases, the exponential term rapidly decays and becomes negligible
as compared to the mass term, so for $(v-\phi)\gg v/\lambda$ the potential
reduces to its quadratic part, $m^2\phi^2/2$. When $\phi$ is larger than $v$,
the interaction term, however, becomes dominant, and the potential falls down
rapidly.  The potential has a local minimum, which at large $\lambda$
is placed
almost exactly at $\phi=0$. This minimum corresponds to the false vacuum,
whereas the true vacuum is $\phi=+\infty$.

The decay of the false vacuum is a non-perturbative phenomenon with the
exponentially small amplitude, which is typical for quantum tunneling
processes. This decay can be described semiclassically by the bounce
configuration, which is a solution to the Euclidean field equation with
finite action, that obeys the boundary condition $\phi(\infty)=0$. The action
of the bounce  determines the exponential suppression factor for
the false vacuum decay,
\[
  \Gamma \sim \e^{-S_0},~~~ S_0\sim v^2.
\]

Since the bounce is  the analog of the instanton, it is suggested in
refs.\cite{Hsu,Voloshin} that theories with unstable vacua may be considered as
models for studying instanton--like scattering. Instead of transitions from
one vacuum to another induced by particle collisions, one studies ``shadow
processes'', where both initial and final states refer to the false vacuum
$\phi=0$  but intermediate states contain a bubble of large positive
field ($\phi>v$ in our case). The simplest shadow process is one described
by the bounce itself, which corresponds
 to the transition at zero energy from the
false vacuum ($\phi=0$ at $\tau=+\infty$) through some bubble--type
configuration ($\phi(x)$ at $\tau=0$) back to the false vacuum
($\phi(x,-\infty)=0$).  We note in passing
 that the very definition of the shadow
processes is of semiclassical nature, and the notion of the shadow processes
may become ambiguous in high energy scattering.

Along with the bounce, there exists another configuration relevant to shadow
processes, namely, the critical bubble, which is a static and unstable
solution to the field equation with finite energy.
The physical significance of the critical bubble is that it is the
minimum static energy configuration on the border, in the configuration
space, between the false and true vacua: being slightly perturbed, the
critical bubble either evolves (in real time) into an expanding domain
of a new phase that eventually fills up the whole space, or shrinks down
to $\phi=0$.
 The critical bubble is the
analog of the sphaleron in gauge theories
(in particular, its free energy determines the rate of the  false
vacuum decay at high enough temperatures).
 In many theories (but not in our
model, see below), the sphaleron mass determines the energy scale where the
instanton--like amplitudes naively become large. For convenience, we will use
the gauge theory terminology and call the critical bubble and bounce by the
sphaleron and instanton, respectively.

\subsection{Sphaleron}

The sphaleron is the static solution to the field equation,
\begin{equation}
  \partial_x^2\phi=m^2\phi-\lambda m^2 v
                   \exp\left[2\lambda\left({\phi\over v}-1\right)\right].
  \label{sph_eq}
\end{equation}
This equation has the same form as the equation of motion for a classical
particle in the upside--down potential $-V(\phi)$ with $x$ playing the role
of time.  The sphaleron corresponds to the motion of the particle from
$\phi=0$ at $x=-\infty$ to $\phi=v$ at $x=0$ where it recoils from the wall
and goes back to $\phi=0$ at $x=\infty$. At large $\lambda$, when the wall is
steep, the recoil occurs in a small interval of ``time'', and most of the
``time'' the particle moves in the region where the nonlinear term in
eq.(\ref{sph_eq}) can be neglected and the potential is quadratic. So,
outside the nonlinearity region, the sphaleron configuration is
\begin{equation}
  \phi=\e^{-m|x|}.
  \label{sph}
\end{equation}
One can check that the region of nonlinearity, where the exponential term in
eq.(\ref{sph_eq}) is comparable with the other terms, is of order
\[
  x\sim{1\over \lambda m},
\]
which is much smaller than  $m^{-1}$. The energy of
the sphaleron,
\[
  E=\int dx \left[{1\over 2}(\partial_x\phi)^2+V(\phi)\right],
\]
is then dominated by the contribution from the region where the field is
linear, $x\gg (\lambda m)^{-1}$; the contribution of the region of
nonlinearity is suppressed by a factor $1/\lambda$. A straightforward
calculation gives
\begin{equation}
  E_{sph}=mv^2
\label{Esph}
\end{equation}
in the leading order in $1/\lambda$.

The fact that the sphaleron mass comes mostly  from the linear region is a
specific feature of our model, which allows, in particular, to
calculate  the momentum distribution and the total number of
particles emitted by a  sphaleron decaying back into the false vacuum.
The field of the decaying sphaleron is linear,
\[
  \phi(x,t)=\int{dk\over\sqrt{2\pi}\sqrt{2\omega_k}}
       (b_k\e^{-i\omega_kt+ikx}+b^*_{-k}\e^{i\omega_kt-ikx}),
\]
and, if the sphaleron begins to decay at $t=0$, obeys the initial conditions
$\phi(x,t=0)=\phi_{sph}(x)$, $\dot{\phi}(x,t=0)=0$. One finds from
eq.(\ref{sph})
\[
  b_k={1\over\sqrt{\pi}}{mv\over\omega_k^{3/2}}.
\]
The total number of particles emitted by the decaying sphaleron is
\begin{equation}
  n_{sph}=\int dk b_k^*b_k={2\over\pi}v^2.
\label{Nsph}
\end{equation}
As expected, the number of particles is of order $v^2$.

\subsection{Instanton}

The instanton is an $O(2)$ symmetric solution to the Euclidean field
equation,
\begin{equation}
  \partial_{\mu}^2\phi=\ddpar{\phi}{r}+{1\over r}\dpar{\phi}{r}=
                      m^2\phi-
      \lambda m^2v\exp\left[2\lambda\left({\phi\over v}-1\right)\right],
  \label{insteq}
\end{equation}
\[
  x^{\mu}=(\tau,x),~~~ r=\sqrt{\tau^2+x^2}.
\]
The appropriate mechanical analog is now a particle moving in the potential
$-V(\phi)$ with the friction coefficient inversely proportional to ``time''
$r$. The motion of the particle begins at some $\phi(0)>v$ with zero velocity
and ends at $\phi=0$ at infinitely large ``time''. The initial potential
energy is lost at $r=\infty$ because of friction.

We are unable to solve eq.(\ref{insteq}) analytically. Nevertheless, by
making the following conjecture on the structure of the field configuration
 we are able to obtain the solution to the leading order
in $1/\lambda$. Namely, in analogy with the sphaleron, we assume that the
size of the region of the Euclidean space--time where the field is nonlinear,
$r_0$, is much smaller than the inverse mass, $r_0\ll m^{-1}$. Outside this
nonlinear ``core'' the field is free, i.e. the nonlinear term in
eq.(\ref{insteq}) is negligible as compared to the kinetic and mass
terms. This outer region $r\gg r_0$ can be in its turn divided into two
regions: $r_0\ll r\ll m^{-1}$ where the mass is unimportant and the field is
both free and massless, and $r~\gtap~m^{-1}$ where the field obeys the
ordinary massive Klein--Gordon equation.  Shortly speaking, we assume that
the field shows different behavior in the following three regions,
\begin{itemize}
\item[(1)] $r~\ltap~r_0$, where
  $\partial_{\mu}^2\phi=
  -\lambda m^2v\exp\left[2\lambda\left({\phi\over v}-1\right)\right]$,
\item[(2)] $r_0 \ll r \ll m^{-1}$, where $\partial_{\mu}^2\phi=0$, and
\item[(3)] $r~\gtap~m^{-1}$, where $\partial_{\mu}^2\phi=m^2\phi$.
\end{itemize}
In the region 2 we will match the two solutions found in the regions
1 and 3 thus obtaining the complete field configuration.  We will justify
our assumptions a posteriori, and now we construct the solution explicitly.

First, consider the regions 1 and 2, where $r\ll m^{-1}$ and the mass
term can be neglected. Since there this no mass scale left over, the equation
is scale invariant. In fact, it is the Liouville equation, which in terms of
the new variable
\begin{equation}
  \chi=-{\lambda\over v}\phi
\label{defX}
\end{equation}
has the form
\begin{equation}
  \partial_{\mu}^2\chi = {4 \over a^2}\e^{-2\chi},
  \label{Liouv_eq}
\end{equation}
where
\begin{equation}
  a={2\e^{\lambda}\over \lambda m}.
  \label{a}
\end{equation}
At the moment we have to consider
the $O(2)$ symmetric solution to the Liouville
equation,
\begin{equation}
  \chi(r)=\ln\left(c+{r^2\over ca^2}\right),
  \label{Liouv_sol}
\end{equation}
where $c$ is an arbitrary constant. Thus in the regions 1 and 2 the field
configuration is
\begin{equation}
  \phi(r)=-{v\over\lambda}\ln\left(c+{r^2\over ca^2}\right),
  \label{phi12}
\end{equation}
where $c$ is yet to be determined. Notice that at $r\gg ca$, where
$\phi(r)\sim\ln r$, the field is free and massless. This means that the
radius $r_0$ separating the regions 1 and 2 is
\begin{equation}
  r_0 = ca.
  \label{r0ca}
\end{equation}

In the regions 2 and 3, i.e. at $r\gg r_0$,
the field is linear, $(\partial^2-m^2)\phi=0$, so the configuration is given
by the $O(2)$ symmetric solution to the Klein--Gordon equation,
\begin{equation}
  \phi(r)=\alpha K_0(mr),
  \label{phi23}
\end{equation}
where $\alpha$ is some constant and $K_0$ is the modified Bessel
function.
 If $r_0\ll m^{-1}$, we can match the two
solutions, eqs.(\ref{phi12}) and (\ref{phi23}), in the region 2. We write
\[
  -{v\over\lambda}\ln\left({r^2\over ca^2}\right)=
  -\alpha\left[\ln{mr\over 2}+\gamma\right],
\]
where we  made use that at small $(mr)$, $K_0(mr)=-\ln(mr/2)-\gamma$, where
$\gamma=0.577\ldots$ is the Euler constant. Taking into account eq.(\ref{a}),
we find the parameters $\alpha$ and $c$,
\[
  \alpha = {2v \over \lambda},
\]
\begin{equation}
  c = \lambda^2\e^{-2(\lambda+\gamma)}.
  \label{alphains}
\end{equation}
So, in the regions 1 and 2, the solution is given by eq.(\ref{phi12}),
and in the regions 2 and 3 it is determined by eq.(\ref{phi23}), with
the
values of $\alpha$ and $c$ fixed by eq.(\ref{alphains}).

It is now straightforward to justify our basic assumption that
$r_0\ll m^{-1}$. We find from eq.(\ref{r0ca})
\begin{equation}
  r_0\sim \lambda\e^{-\lambda}{1\over m}
  \label{coresize}
\end{equation}
which is smaller than $m^{-1}$ by an exponential factor $\e^{-\lambda}$. So
our basic assumptions about the structure of the field are indeed valid with
exponential accuracy.

Thus, we have found the instanton by solving separately the field equation in
the regions of small and large $r$ and matching the two solutions at
intermediate $r$. The assumptions made on the structure of the instanton
solution are self-consistent at large $\lambda$. The field configuration
consists of an exponentially small core, where $\phi$ is nonlinear, and a
large tail that extends from $r_0\sim\e^{-\lambda}m^{-1}$ to $m^{-1}$.
In fact, it is straightforward to see that in the whole Euclidean
 space--time our
solution satisfies the field equation with the accuracy  $\e^{-\lambda}$.

Note that the value of the field $\phi$ at the center of the instanton is
finite in the limit $\lambda\to\infty$,
\begin{equation}
  \phi(0)=-{v\over\lambda}\ln c = 2v.
\label{center}
\end{equation}

The instanton action is dominated by the contribution of the kinetic term in
the lagrangian, and comes mostly from the region 2,
\begin{equation}
  S_0=2\pi\int rdr \half (\partial_r\phi)^2=
      {4\pi v^2\over\lambda}\left(1-{\ln\lambda\over\lambda}
             +O(\lambda^{-1})\right).
  \label{inst_act}
\end{equation}
Recall that we take $v\gg\lambda\gg 1$, so the   vacuum decay
probability is
exponentially small.

\subsection{Leading order instanton approximation}

In the leading order of the perturbation theory around the instanton, the
two--particle cross section at energy $E$ can be found by the technique of
ref.\cite{KRT}. One writes for the exponent of eq.(\ref{FE}),
\begin{equation}
  F(E)=2S_0-ET-\int dk R^*(k)R(k)\e^{-\omega_kT},
  \label{FR}
\end{equation}
where $R(k)$ is the Fourier component of the asymptotics of the instanton
field,
\begin{equation}
  \phi(x,\tau)=\int{dk\over\sqrt{2\pi}\sqrt{2\omega_k}}
               R(k)\e^{-\omega_k\tau+ikx},~~~\tau\to\infty,
  \label{phiR}
\end{equation}
and $T$ is related to the energy by the following condition,
\begin{equation}
  E=\int dk \omega_k R^*(k)R(k)\e^{-\omega_kT}.
  \label{E_R}
\end{equation}
The Fourier components in eq.(\ref{FR}) can be calculated from
eqs.(\ref{phi23}) and (\ref{alphains}),
\[
  R(k)=\sqrt{\pi\over\omega_k}{2v\over\lambda}.
\]
The integrals in eqs.(\ref{FR}) and (\ref{E_R}) are then straightforward to
evaluate, and for $E\gg E_{sph}$ one obtains the leading order
instanton approximation result,
\begin{equation}
  F(E)= 2S_0\left(1-{1\over\lambda}\ln{E\over E_{sph}}+
        O\left({\ln\lambda\over\lambda}\right)\right).
  \label{ring}
\end{equation}
Naively, eq.(\ref{ring}) implies that the cross section becomes large at
\begin{equation}
  E\sim E_0=\e^{\lambda}E_{sph}.
  \label{E_Ring}
\end{equation}
This scale is much higher than $E_{sph}$, which is a peculiar feature of our
model.

\section{Improved dilute instanton gas approximation}

\subsection{Difficulties of the ordinary dilute instanton gas}

At low energies, the classical boundary value problem, eqs.(\ref{bvp1})
and (\ref{bvp2}), can be solved by making use of
the dilute instanton gas approximation \cite{RST,KRTperiod}.
Typically, the dilute instanton gas approximation breaks down at $E\sim
E_{sph}$;
this is the case, for example, in the (1+1)-dimensional Abelian Higgs
model \cite{RST} and the electroweak theory \cite{KRTperiod}. Let us see,
however, that our model is peculiar in this respect: the ordinary dilute
instanton gas approximation breaks down at energies much smaller than
$E_{sph}$.

Let us consider, for example, the boundary value problem at $\theta=0$.
In that case the solution is periodic in Euclidean time with the period
$T$ \cite{KRTperiod}. The basic assumption of the dilute instanton gas
approximation is that the Euclidean solution may be represented by the
sum
of widely separated instantons, which, to the leading order, do not
distort each other's interiors. Thus, the periodic instanton field is
approximated by the sum
\begin{equation}
  \phi(x) = \sum_l\phi_{inst}(x-x_l),
  \label{sumphi}
\end{equation}
where, by periodicity, $x_l=(lT,0)$ are the positions
of the instantons in the chain (there are no
anti-instantons in our model).

At  first sight, the dilute instanton gas would be a good
approximation if $T$ is much larger than the core size $r_0$ given by
eq.(\ref{coresize}). However, we will see immediately that this
condition
is not sufficient for a two-fold reason. First, the field in the center
of the instanton, eq.(\ref{center}), is strong enough, so that the
exponential interaction is operative. Therefore, even relatively small
distortion of the field due to other instantons may have strong effect
on the instanton interior. Second, the field of each instanton falls off
rather slowly at $|x-x_l|\ll m^{-1}$, so that neighboring instantons
may produce a collective effect on a given instanton.

To establish the region of validity of the dilute gas approximation, let
us consider the interior region of an instanton sitting at $x=0$ and
write explicitly  its contribution to the sum in
eq.({\ref{sumphi}),
\begin{equation}
  \phi(x)=\phi_{inst}(x)+\tilde{\phi}(x),
  \label{tilde}
\end{equation}
where
\[
 \tilde{\phi}(x)=\sum_{l\neq 0}\phi_{inst}(x-x_l)
\]
is the distortion of the field due to other instantons.
For the dilute instanton gas to work, the distortion
$\tilde{\phi}$ should not change the field equation considerably.
In particular, at $x=0$ one should have
\[
  \exp\left[2\lambda\left({\phi_{inst}+\tilde{\phi}\over v}-1\right)
  \right]\approx\exp\left[2\lambda\left({\phi_{inst}\over v}-1\right)
  \right],
\]
which means
\begin{equation}
  \tilde{\phi}(x=0)\ll {v\over\lambda}.
\label{JUNK}
\end{equation}
Recalling that, due to eqs.(\ref{phi23}) and (\ref{alphains}),
 \[
\phi(x=0)=\sum_{l\neq 0}\frac{2v}{\lambda}K_0(m|l|T),
\]
one finds that eq.(\ref{JUNK}) is valid only when
\[
  T\gg m^{-1}.
\]
Obviously, this condition is much stronger than the naive estimate
$T\gg r_0$. We will see in Sect 3.2 that the corresponding condition
on energy is
\[
  E\ll {E_{sph}\over \lambda^2},
\]
so that the ordinary dilute instanton gas approximation breaks down
at energies much lower
than the sphaleron mass.

\subsection{Improved dilute gas}

The ordinary dilute instanton gas approximation breaks down long before
the instanton cores begin to overlap. One may expect, however, that the
picture of well separated instanton-like objects is valid in a wider
region of energies, but each instanton is strongly affected by the
collective field of the others. When the instanton core is much smaller
than the separation between the instantons, this collective field is
constant in the instanton interior, so the distorted instanton is still
$O(2)$ symmetric. Furthermore, outside the instanton cores, the field is
still linear, so that we are still able to make use of the assumptions
described in Sect.2.3.

To obtain the solution to the boundary value problem, eqs.(\ref{bvp1}) and
(\ref{bvp2}), we begin with the region outside the cores. Let the
distorted instantons be located at $x=0$ and
$\tau=(\pm T_1\pm lT, 0)$, $l=0,1,2\ldots$, where $T_1$ is yet unknown
parameter (see fig.3).
At $\theta\neq 0$, only two of these instantons
(with $l=0$) have unit strength, while others are fake instantons
with strength $\exp(-|l|\theta)$ (cf. ref.\cite{RST}). Recall that we are
interested in the field on the contour ACD in the
complex time plane shown in
fig.3, so we are able to  ignore interiors of all instantons except for
one sitting at $\tau=T_1$.

  Thus, the field outside the core is
\begin{equation}
  \phi(x,\tau) = \alpha\sum_{l=-\infty}^{l=\infty}\e^{-|l|\theta}
         \left[K_0\left(m\sqrt{(\tau-T_1-lT)^2+x^2}\right)
         +K_0\left(m\sqrt{(\tau+T_1-lT)^2+x^2}\right)\right],
  \label{dilgas}
\end{equation}
where $\alpha$ is yet unknown  constant that may be different from
eq.(\ref{alphains}).
The field (\ref{dilgas}) automatically
satisfies the boundary conditions, eq.(\ref{bvp2}). The Fourier components
of the asymptotics of the field in the initial and final states are
\[
  f_k =f^*_k= \alpha\sqrt{\pi\over\omega_k}{2\e^{-\omega_kT/2-\theta}
        \cosh(\omega_kT_1) \over 1-\e^{-\omega_kT-\theta}},
\]
\begin{equation}
  b_k=b_k^*=\alpha\sqrt{\pi\over\omega_k}{\e^{-\omega_kT_1}+
            \e^{-\omega_k(T-T_1)-\theta}\over 1-\e^{\omega_kT-\theta}}.
  \label{Fourier}
\end{equation}
The energy $E$ and the number of initial particles $n$ are related to the
parameters  $T$, $T_1$ and $\theta$ as follows,
\[
  E = \int dk \omega_k\e^{\theta}f_k^* f_k = \pi\alpha^2\int dk
      {\e^{-\omega T-\theta}(2\cosh\omega_k T_1)^2\over
      (1-\e^{-\omega_kT-\theta})^2}
\]
\begin{equation}
    = \int dk \omega_k b_k^* b_k = \pi\alpha^2 \int dk
      {(\e^{-\omega_kT_1}+\e^{-\omega_k(T-T_1)-\theta})^2\over
      (1-\e^{-\omega_kT-\theta})^2},
  \label{dil_E}
\end{equation}
\begin{equation}
  n = \int dk \e^{-\theta}f_k^* f_k = \pi\alpha^2\int {dk\over \omega_k}
      {\e^{-\omega T-\theta}(2\cosh\omega_k T_1)^2\over
      (1-\e^{-\omega_kT-\theta})^2}.
  \label{dil_n}
\end{equation}
In fact, these three relations are consequences of eq.(\ref{derivS})
and the condition of equilibrium of forces acting on the instanton
sitting at $\tau=T_1$. Eqs.(\ref{dil_E}) and (\ref{dil_n}) enable one to
express $T$, $T_1$ and $\theta$ through the energy and number of initial
particles, once the value of $\alpha$ is known. Alternatively, one may
consider the boundary value problem at given values of $T$ and $\theta$,
and relate these parameters to $E$ and $n$ at the very last step. In
that case eq.(\ref{dil_E}) should be used to express $T_1$ in terms of
$T$ and $\theta$.

To find the configuration inside the core, let us first
consider the region close to the  instanton located at $(T_1,0)$,
but still outside the core. In this region, all  terms in the sum
(\ref{dilgas}) are constant with the accuracy  $r_0/T$, except for one
that refers to the instanton $(T_1,0)$. Eq.(\ref{dilgas}) then
reduces to a form that is $O(2)$ symmetric with respect to the point
$(T_1,0)$,
\begin{equation}
  \phi(x,\tau)= \alpha\left[K_0\left(m\sqrt{(\tau-T_1)^2+x^2}\right)+f\right],
  \label{outcore}
\end{equation}
where $f$ is a constant that depends on $T$ and $\theta$,
\begin{equation}
  f=2\sum_{l=1}^{\infty}\e^{-l\theta}K(nmT)+
    \sum_{l=-\infty}^{\infty}\e^{-|l|\theta}K_0(m|lT-2T_1|).
\label{15X}
\end{equation}
A convenient representation for this constant is
\begin{equation}
   f= \int {dk \over 2\omega_k}
      {2\e^{-\omega_kT-\theta}+\e^{-\omega_kT-\theta+2\omega_kT_1}+
      \e^{-2\omega_kT_1} \over 1-e^{-\omega_kT-\theta}}.
  \label{f}
\end{equation}

Eq.(\ref{outcore}) can be considered as the boundary condition for
the Liouville equation that determines the field inside the core. Since this
boundary condition is $O(2)$ symmetric, the field inside the core is still
given by the $O(2)$ symmetric solution of the Liouville equation
(\ref{Liouv_eq}),
\begin{equation}
  \phi(r)=-{v\over\lambda}\ln\left(c+{(\tau-T_1)^2+x^2\over ca^2}\right),
  \label{incore}
\end{equation}
where $a$ is given by eq.(\ref{a}) and
$c$ is some constant that can differ from that of the single instanton,
eq.(\ref{alphains}). If the core size, which is $r_0=ca$, is much smaller
than both $T_1$ and $T/2-T_1$, one can match eqs.(\ref{outcore})
and (\ref{incore}) in the region $r_0^2\ll (\tau-T_1)^2+x^2\ll
T_1^2$, $(T/2-T_1)^2$ where $\phi$ is free and massless. In this way we obtain
the values of the parameters $c$ and $\alpha$,
\begin{equation}
  c=\lambda^2\e^{-2(\lambda+\gamma-f)},
  \label{cf}
\end{equation}
\begin{equation}
  \alpha={2v\over\lambda}.
  \label{alpha}
\end{equation}

Eqs.(\ref{outcore}) -- (\ref{alpha}) determine
the solution to the boundary value problem. For given $T$ and $\theta$,
the value of $T_1$ is determined by eq.(\ref{dil_E}). The value of $f$ is
then given by eq.(\ref{f}), so that all parameters become known.
Outside the
cores, the field is exactly the same as  in the ordinary dilute
instanton gas approximation: the configuration (\ref{dilgas}) with
$\alpha$
given by eq.(\ref{alpha}) is precisely the sum of undistorted instanton
fields. On the other hand,
the size of the core and the field inside the core are modified.
Namely, the core size is
\begin{equation}
  r_0=ca=2\lambda \exp(-\lambda+2f-2\gamma){1\over m},
  \label{r0f}
\end{equation}
while the field configuration inside the core is  the $O(2)$
symmetric solution of the Liouville equation with the parameter $c$
given by
eq.(\ref{cf}).

We will call this procedure the ``improved dilute
instanton gas approximation''.  This approximation is valid provided the
core size, eq.(\ref{r0f}), is much smaller than $T_1$ and
$T/2-T_1$,
\[
  \lambda \exp(-\lambda+2f){1\over m}\ll T,~(T/2-T_1).
\]
The explicit formulas  will be given in
Sects. 3.3 and 3.4 where we will also discuss the actual region of
validity of the improved dilute instanton gas approximation. We note in
passing that at $T\gg m^{-1}$, the improved and ordinary dilute gas
approximations coincide. Indeed, in that case all
terms on the right hand side of eq.(\ref{15X}) are much smaller than 1,
so that $f\ll 1$. The value of the parameter $c$, eq.(\ref{cf}), is then the
 same as for an isolated instanton, eq.(\ref{alphains}), and the field
 in the instanton interior is precisely the field of an isolated instanton.

When $\theta$ is fixed, and $T$ decreases, the value of $f$ increases. Since
$f$ enters  the expression for $r_0$, eq.(\ref{r0f}), in the
exponent, the core size increases rapidly when $T$ falls below $m^{-1}$.  At
some $T=T_{crit}$ which depends on $\theta$, the core size becomes
comparable to the distance between the instantons and our approximation
breaks down. We will see that this happens at energies comparable to,
 or even much
larger than the sphaleron mass, depending on the value of $\theta$.

Now let us evaluate the action for our configuration. Since the core is
exponentially small, the action comes entirely from the region where the
field is linear, so one has
(recall that we choose the convention that the action is real and
positive for real Euclidean fields)
\begin{equation}
  S=-{i\over 2}\int dtdx((\partial_{\mu}\phi)^2-m^2\phi^2),
  \label{Sint}
\end{equation}
where the integration over time is performed along the contour of fig.3.
This contour can be divided into two parts: $(iT/2-\infty,iT_1)$, and
$(iT_1,+\infty)$. In the first part the field is given
by the initial state asymptotics, while in the second part
it is given by the final state asymptotics. Substituting these asymptotics,
eq.(\ref{bvp2}), into the integral (\ref{Sint}), one obtains
\[
  S={1\over 4}\int dk (f^*_kf^*_{-k}e^{\omega_k(T-2T_1)+2\theta}
     -f_kf_{-k}\e^{-\omega_k(T-2T_1)})
\]
\[
   -{1\over 4}\int dk (b^*_kb^*_{-k}e^{-2\omega_kT_1}
     -b_kb_{-k}\e^{2\omega_kT_1)}).
\]
The explicit form of the Fourier components is given by
eq.(\ref{Fourier}), so we find
\[
  S={4\pi v^2\over\lambda^2} \int {dk \over 2\omega_k}
      {1+\e^{-\omega_kT-\theta}+\e^{-\omega_kT-\theta+2\omega_kT_1} +
      \e^{-2\omega_kT_1} \over 1- e^{-\omega_kT-\theta}}.
\]
Comparing this expression with eq.(\ref{f}), we obtain
\[
  S={4\pi v^2\over\lambda^2}\left(\int {dk \over 2\omega_k}+f\right).
\]
The integral in this equation diverges logarithmically in the ultraviolet
region. However, we have to cut  this integral off at the scale $1/r_0$, since
eq.(\ref{Fourier}) is valid only at momenta much lower than $1/r_0$.
So, we have
\[
  S={4\pi v^2\over\lambda^2}\left(\ln{1\over mr_0}+f\right).
\]
Recalling that the core size $r_0$ is related to $f$ by eq.(\ref{r0f}),
we obtain finally
\begin{equation}
  S={4\pi  v^2\over\lambda}\left(1-{f\over\lambda}+
     O\left( {\ln\lambda\over\lambda}\right)
  \right)
  \label{S}
\end{equation}
It is straightforward to verify that this action as function of $T$ and
$\theta$
satisfies eq.(\ref{derivS}).

To obtain the explicit formulas for the transition probabilities at a
given
number of initial particles, we have to consider various limiting cases.
Since there exists an extra large parameter $\lambda$ besides the
``coupling constant'' $v$, we distinguish the cases
$\nu=n/v^2\sim 1$ and $\nu\sim 1/\lambda$ where $n$ is
the number of the initial particles. We will see in what follows that
these cases correspond to $\theta\sim 1/\lambda$ and
$\theta\sim 1$, respectively. Let us discuss the two regimes in turn.

\subsection{$\theta\sim 1/\lambda$}

Let us first consider the case of small $\theta$,
$\theta\sim 1/\lambda$.
At small
$\theta$, the condition of energy conservation, eq.(\ref{dil_E}), implies that
$T_1\approx T/4$.
Let us also  assume that $T\sim (\lambda m)^{-1}$; we will see that this
regime corresponds to $E\sim E_{sph}$.
For these values of $T$ and $\theta$, the number of relevant terms in
eq.(\ref{15X})  is of order $\lambda$, which is large. In other words,
there are many fake instantons that determine the solution on the
contour of fig.3. For the actual
calculation of $f$, it is more convenient to make use of the integral
representation, eq.(\ref{f}). At $k\ll\lambda m$, the
integrand in eq.(\ref{f}) can be expanded in the following way,
\[
    {2\e^{-\omega_kT-\theta}+\e^{-\omega_kT-\theta+2\omega_kT_1}
    +\e^{-2\omega_kT_1} \over 1 - \e^{-\omega_kT-\theta}}=
  {4\over\omega_kT+\theta}-1+O(\lambda^{-1}).
\]
So, one has
\[
  f=\int{dk\over 2\omega_k}{4\over\omega_kT+\theta}-\int{dk\over 2\omega_k}.
\]
The second integral on the right side diverges logarithmically in the
ultraviolet, and should be
cut--off at $k\sim\lambda m$.
We obtain
\begin{equation}
  f=\frac{2\pi}{mT}f_1\left(\theta/mT\right)-\ln\lambda+O(1),
  \label{fbeta}
\end{equation}
where
\[
  f_1(\beta) ={2\over\pi}{\arccosh\beta\over\sqrt{\beta^2-1}}.
\]
Note that by our assumptions, $\theta\sim 1/\lambda$,
 $mT\sim1/\lambda$, the first term in eq.(\ref{fbeta})
 is of order $\lambda$.
Analogously,  the energy and the number of initial particles
are obtained from eqs.(\ref{dil_E}) and (\ref{dil_n}),
\begin{equation}
  E=\left(\frac{4\pi}{\lambda
  mT}\right)^2f_2\left(\theta/mT\right)E_{sph},~~~
n=\left(\frac{4\pi}{\lambda mT}\right)^2f_3\left(\theta/mT\right)n_{sph},
  \label{Enbeta}
\end{equation}

where
\[
  f_2(\beta) ={2\over\pi}\left({\beta\over\beta^2-1}-
         {\arccosh\beta\over(\beta^2-1)^{3/2}}\right),
\]
\[
  f_3(\beta) ={\pi\over 2\beta}(f_1(\beta)-f_2(\beta)),
\]
and the energy and the number of particles for the sphaleron are given
by eqs.(\ref{Esph}) and (\ref{Nsph}).
We see that the regime $\theta\sim 1/\lambda$, $mT\sim
1/\lambda$ occurs when the
energy $E$ and the number of particles, $n$, are of the same order of
magnitude as the corresponding quantities for the sphaleron, $E_{sph}$ and
$n_{sph}$. Since $n_{sph}\sim v^2$, we conclude that the regime
$\theta\sim 1/\lambda$ corresponds to $\nu=n/v^2\sim 1$.

The exponent for the total probability is then straightforward to evaluate,
\begin{equation}
  F={8\pi v^2\over\lambda}\left(1-\frac{4\pi}{\lambda mT}
    f_1\left(\theta/mT\right)\right).
  \label{Fbeta}
\end{equation}
The condition for the validity of the improved dilute instanton gas
approximation is $r_0\ll T$. Since $T\sim(\lambda m)^{-1}$, and
\[
  r_0\sim\lambda\exp(-\lambda+2f){1\over m},
\]
the approximation is valid if
\begin{equation}
  1-\frac{4\pi}{\lambda mT}f_1\left(\theta/mT\right)\gg{1\over\lambda}.
  \label{valid}
\end{equation}
Comparing eqs.(\ref{Fbeta}) and (\ref{valid}) we see that we can trust
the dilute gas approximation up until the suppression
becomes much weaker than the instanton suppression,
\[
  F\sim\frac{S_0}{\lambda}.
\]

In the particular case $\theta=0$, the field configuration is the
periodic instanton that describes the most probable process at
given energy $E$ \cite{KRTperiod} (the maximization over all possible
values of the number of initial particles $n$ can be seen from the relation
$\dpar{F}{n}=-\theta/v^2=0$). Since the  functions
$f_1$, $f_2$, $f_3$ are normalized in such a way  that
$f_1(0)=f_2(0)=f_3(0)=1$, we
have for the periodic instanton
\[
 {E\over E_{sph}}=\left(\frac{4\pi}{\lambda mT}\right)^2,
\]
 and the function $F(E)$ can be found explicitly,
\begin{equation}
  F(E)={8\pi v^2\over\lambda}\left(1-\left({E\over
E_{sph}}\right)^{1/2}\right).
  \label{FEperiod}
\end{equation}
The exponent for the probability, $(-F(E))$ for the periodic instanton
is shown in
fig.4. It increases from $-2S_0$ at low energies to zero at $E=E_{sph}$.
The number of particles for the periodic instanton increases with
energy,
\[
  n=\frac{E}{E_{sph}} n_{sph},
\]
so that the average energy per particle remains constant.
The core size of an individual instanton rapidly increases with energy,
\[
r_0\sim\frac{1}{\lambda
m}\exp\left[-\lambda\left(1-\left(\frac{E}{E_{sph}}\right)^{1\over
2}\right)\right].
\]

Strictly speaking, there is an energy
 region close to the sphaleron mass where the
improved dilute instanton gas
approximation does not work. According to eq.(\ref{valid}), this region is
\[
  1-{E\over E_{sph}}\sim {1\over\lambda}.
\]
We will discuss this region in detail in Sect.4 and show that the
behavior
(\ref{FEperiod}) indeed persists up to $E_{sph}$.

Let us now consider the general case $\theta\neq 0$, still assuming
$\theta\sim 1/\lambda$.
To find the probability of the transition at fixed energy $E$ and number of
initial particles $n$, which are of the order of  $E_{sph}$ and
$n_{sph}$, respectively, one finds $T$ and $\theta$ from eq.(\ref{Enbeta})
and then substitutes them into eq.(\ref{Fbeta}). The
result is presented in fig.5. For a given $n<n_{sph}$ the
function $F(E)$ starts from $E=\pi n m/2$, where $\theta=0$ and the
configuration is just the periodic instanton, and reaches zero
at some $E_{crit}$
determined by the condition
\begin{equation}
 1-\frac{4\pi}{\lambda mT}f_1\left(\theta/mT\right)=0.
\label{20X}
\end{equation}
Very close to $E_{crit}$, however,
 the improved dilute instanton gas approximation is not
reliable, see eq.(\ref{valid}).

 In the whole energy interval,
$F$ is a monotonically increasing function of energy. As in the case of the
periodic instanton, at the  point where the approximation breaks down,
the function $F$ is of order $S_0/\lambda$. The critical energy at which
 $F$ becomes small, is a function of the number of initial
particles: the smaller the number of initial particles, the larger the
critical energy.

At $1/\lambda\ll n/ n_{sph}\ll 1$, the critical energy can be calculated
analytically.
In that case one has $\theta/mT\gg 1$, so that one makes use of the
asymptotics of $f_1$, $f_2$ and $f_3$ at large values of their argument,
\[
  f_1(\beta)={2\over\pi}\frac{\ln\beta}{\beta},
\]
\[
  f_2(\beta)={2\over\pi}{1\over\beta},
\]
\[
f_3(\beta)=\frac{\ln\beta}{\beta^2}.
\]
It is then straightforward to express $T$ and $\theta$ through $E$ and $n$
from eq.(\ref{Enbeta}) and then obtain the critical energy at which
eq.(\ref{20X}) is satisfied
and the instanton suppression is strongly reduced.
 One finds the following exponential dependence,
\[
 \frac{E_{crit}(n)}{E_{sph}}\sim\exp\left({\pi^2\over 4}
       {n_{sph}\over n}\right).
\]
Note that we consider $n/n_{sph}\gg 1/\lambda$, so $E_{crit}$ is still
much smaller than the energy scale where the two--particle amplitudes
naively become large, eq.(\ref{E_Ring}).

\subsection{$\theta\sim 1$}

In the case $\theta\sim 1$, the sum in eq.(\ref{15X}) is saturated by a finite
number of terms. Let us consider the case $mT\ll 1$, which is, of
course, of primary interest.  Then one can replace the modified Bessel
functions in eq.(\ref{15X}) by their values at small argument,
$-\ln(mT)+O(1)$, so, to the logarithmic accuracy, one has
\begin{equation}
  f = {1+3\e^{-\theta}\over 1-\e^{-\theta}}\ln{1\over mT}.
  \label{fkappa}
\end{equation}
The action, eq.(\ref{S}), becomes
\[
  S=S_0\left(1-{1\over\lambda}{1+3\e^{-\theta}\over 1-\e^{-\theta}}
    \ln{1\over mT}\right).
\]
We see that the parameter $T$ enters into the action in the combination
$\lambda^{-1}\ln(1/mT)$, so in this case it is natural to consider
exponentially small $T$.
 Let us introduce a new parameter, instead of $T$,
\begin{equation}
  \kappa={1\over\lambda}\ln{1\over mT},
  \label{Tkappa}
\end{equation}
and  assume that
\[
\kappa\sim 1.
\]
 These values of $T$ correspond to exponentially large
energies: indeed, from the relation
\[
  E=\frac{\partial S}{\partial T}\sim \frac{v^2}{\lambda^2 T}
\]
we find that
\begin{equation}
 \frac{E}{ E_{sph}}\sim\e^{\lambda\kappa}.
  \label{Ekappa}
\end{equation}
Analogously, the number of initial particles is
\begin{equation}
  n={16\pi^2\over\lambda}{\e^{-\theta}\over (1-e^{-\theta})^2}\kappa n_{sph}.
  \label{nkappa}
\end{equation}
Since both $\theta$ and $\kappa$ are of order 1, eq.(\ref{nkappa})
implies that the number of initial particles is
small,
\[
   \frac{n}{n_{sph}}\sim\frac{1}{\lambda}.
\]

To calculate the function $F$, one notices that $ET$ is negligible as
compared to $S$. So one has
\begin{equation}
  F=2S-n\theta=2S_0\left[1-\left({1+3\e^{-\theta}\over 1-\e^{-\theta}}+
    {4\theta\e^{-\theta}\over (1-\e^{-\theta})^2}\right)\kappa\right],
    \label{Fkappa}
\end{equation}
where $\theta$ and $\kappa$ as functions of $E$ and $n$ are determined
by eqs.(\ref{Ekappa}) and (\ref{nkappa}). Explicitly,
\[
  \theta=\arccosh\left(1+{8\pi^2\over\lambda^2}{n_{sph}\over n}
         \ln{E\over E_{sph}}\right),
\]
and
\[
  \kappa=\frac{1}{\lambda}\ln{E\over E_{sph}}.
\]

According to eqs.(\ref{r0f}) and (\ref{fkappa}), the size of the
individual instanton increases with energy,
\[
r_0\propto\exp\left[ -\lambda\left(1 -
2\frac{1+3\e^{-\theta}}{1-\e^{-\theta}}\kappa\right)\right],
\]
while the separation between the instantons decreases,
\[
T\propto\exp(-\lambda\kappa).
\]
The  validity of the improved dilute instanton gas approximation
becomes suddenly lost at the energy $E_{crit}$ when
\begin{equation}
  \kappa=\kappa_{crit}=\frac{1-\e^{-\theta}}{3+5\e^{-\theta}}.
  \label{kappacrit}
\end{equation}
It follows from
eq.(\ref{Fkappa}) that at the critical energy, the
function $F$ is smaller than $2S_0$ only by a factor of order 1,  contrary
to the case $\theta\sim 1/\lambda$ where the reduction of the function $F$
is of order $\lambda^{-1}$. Thus, at small number of initial particles,
$n/n_{sph}\sim 1/\lambda$, our approximation breaks down when the
probability is still exponentially small, with the exponent of order of,
but smaller than, the instanton one.

The above formulas simplify considerably in the most interesting case
$n/n_{sph}\to 0$, which occurs in the limit $\theta\to \infty$.
 It is clear from eq.(\ref{Fkappa}) that this limit is smooth,
and the exponent for the probability is
\begin{equation}
  F(E)=2S_0\left(1-\frac{1}{\lambda}\ln{E\over E_{sph}}\right).
\label{IX}
\end{equation}
Surprisingly enough, eq.(\ref{IX}) coincides with the leading order
instanton result, eq.(\ref{ring}).

Eq.(\ref{IX})
is reliable at $E<E_{crit}$ where $E_{crit}$ is determined from
eq.(\ref{kappacrit}) to be
\begin{equation}
  E_{crit}\sim\e^{\lambda/3}E_{sph}.
\label{critE}
\end{equation}
This energy is exponentially large in $\lambda$, but still much smaller
than the scale $E_0\sim\e^{\lambda}E_{sph}$ set by the leading order
calculation. The exponent of the probability at $E\sim E_{crit}$ is
numerically smaller than the instanton action,
\[
  F_{crit}={2\over 3}(2S_0),
\]
but the probability is still suppressed exponentially.

Thus, in the most interesting case of small number of initial particles,
the improved dilute instanton gas approximation enables one to go beyond
the perturbation theory about the instanton, but not very far beyond.

\subsection{Multi--instanton amplitudes}

The interest in multi--instanton processes in the context of baryon number
violation has been raised by the conjecture \cite{Zakharov1,M&S}
 that the
multi--instanton contributions may become essential at energies where the
one--instanton amplitude is still small. This conjecture is based on the
observation that since the $many\to many$ amplitude becomes large at the
sphaleron mass, the transition from an initial few--particle state
that goes through a chain of $many\to many$ processes (fig.6)
in principle may become comparable with the one--instanton one at energies
larger than the sphaleron mass.
This conjecture, that has lead to the premature unitarization hypothesis, is
an object of controversial discussion.

In our model, the function $F$ is reliably calculable at energies above
$E_{sph}$.
Furthermore, as discussed in Sect.3.3, the case
when $n$ is of order of, but smaller than $n_{sph}$  can be treated
up to the energy where $F$ becomes parametrically smaller than $S_0$.
Thus, our model provides a means to test  the relevance of
multi-instantons above the sphaleron energy.
Let us
 see that the multi--instanton amplitudes are  not essential when
the one--instanton one is small, and, moreover, the $p$--instanton amplitude is
exponentially smaller than the $(p-1)$--instanton one, even at energies larger
than
the sphaleron mass. This will provide a counterexample to the claim
 of refs.\cite{Zakharov1,M&S}.

Consider $p$--instanton processes at $n$ of order of, but smaller than
$n_{sph}$.
 The field configuration now consists of
$2p$ instantons located at $\pm T_j$, $j=1,2,\ldots,p$ and an infinite set
of fake small instantons at $\pm T_j\pm lT$ with intensities
$\e^{-|l|\theta}$ (the case $p=2$ is shown in fig.7).
For $\theta\ll 1$, the distances between the instantons are approximately
equal, so
\[
  T_j\approx {2j-1\over 4p}T.
\]
The value of  the parameter $f$ is now
\[
  f=2\sum_{l=1}^{\infty}\e^{-l\theta}K_0(lmT)+
    \sum_{j=2}^p\sum_{l=-\infty}^{\infty}\e^{-|l|\theta}K_0(m|lT+T_j-T_1|)
\]
\[
    +\sum_{j=1}^p\sum_{l=-\infty}^{\infty}\e^{-|l|\theta}K_0(m|lT+T_j+T_1|)
  =p\frac{2\pi}{mT_p}f_1\left(\theta_p/mT_p\right)-\ln\lambda+O(1),
\]
where $T_p$ and $\theta_p$ may differ from their one-instanton values
(it is energy and number of initial particles that are fixed).
In complete analogy to Sect.3.3, one finds for
the energy and  number of initial particles
\begin{equation}
  E=p^2\left(\frac{4\pi}{\lambda
  mT_p}\right)^2f_2\left(\theta_p/mT_p\right),
{}~~~n=p^2\left(\frac{4\pi}{\lambda mT_p}\right)^2
     f_3\left(\theta_p/ mT_p\right),
  \label{pEn}
\end{equation}
while the exponential suppression function is
\begin{equation}
  F_p=2pS_0\left(1-p\frac{4\pi}{\lambda mT_p}
  f_1\left(\theta_p/mT_p\right)\right).
  \label{multiF}
\end{equation}
To calculate the function $F_p(E,n)$ one solves eq.(\ref{pEn})
with respect to $\theta_p$ and $T_p$ and then substitutes their values into
eq.(\ref{multiF}). From eq.(\ref{pEn}) it is clear that for given
 $E$ and $n$, the values of $T_p$ and $\theta_p$ are
$p$ times larger than the their values in the one--instanton case.
So, the $p$--instanton function $F_p(E,n)$ is $p$ times larger than the
one--instanton suppression function at the same energy and  number of
initial particles,
\[
  F_p(E,n)=pF(E,n).
\]
We conclude that at those energies when the one--instanton
 amplitude is exponentially suppressed, all
multi--instanton contributions are suppressed even stronger.
 For a given
number of initial particles, the suppression of multi--instantons
 disappears
at the same energy as the suppression of the one--instanton contribution.

\section{Beyond the improved dilute instanton gas approximation:
          periodic instanton}

The improved dilute instanton gas approximation breaks down when the
size of the core, $r_0$, becomes of order of the distance between the
instantons, $T$, so that the instanton cores begin to overlap.
At this energy, however, both $r_0$ and $T$ are much smaller than
$1/m$,
so one may hope to calculate the exponent for the transition probability
in the leading order in $1/\lambda$ beyond the improved dilute gas.
Indeed, one may expect  the exponential term in the lagrangian to play a
role only in a small region of space-time of order $r_0$ or $T$; in that
region the mass term may be neglected, and the field obeys the Liouville
equation.
Outside the non-linearity region, but at distances much smaller than
$1/m$,
the field is both free and massless, so that the matching of the
solutions to the Liouville equation and massive Klein-Gordon equation
may still be possible.

It is almost obvious that this procedure should work for periodic
instantons at all energies up to $E_{sph}$: both the improved instanton
and the sphaleron have cores whose sizes are indeed smaller than $1/m$.
In this section we construct the periodic instantons explicitly in the
region of energies where $(E_{sph}-E)/E_{sph}\sim 1/\lambda$, and the
improved
dilute gas approximation does not work.

Let us first discuss the region of non-linearity where the field obeys
the Liouville equation. We do not expect that the solution is $O(2)$
symmetric (obviously, the sphaleron is  not), so we have to consider a
general solution
to the Liouville equation,
\begin{equation}
  \chi(z,\bar{z})=\frac{1}{2}\ln
        \frac{\left(1+f(z)g(\bar{z})\right)^2}{f'(z)g'(\bar{z})},
  \label{gensol}
\end{equation}
where the variable $\chi$ is introduced in eq.(\ref{defX}),
\begin{equation}
  z={x+i\tau\over a},~~~ \bar{z}={x-i\tau\over a}.
  \label{compvar}
\end{equation}
$a$ is defined in eq.(\ref{a}) and $f(z)$ and $g(\bar{z})$ are arbitrary
functions.

To specify the solution that describes the periodic instanton, we point
out first that the ordinary instanton is determined by the functions
\begin{equation}
  f(z)={c\over z},~~~g(\bar{z})={c\over\bar{z}},
  \label{fginst}
\end{equation}
that have poles at the position of the instanton. So, we impose the
following requirements on the
functions relevant for the periodic instanton:

i) periodicity,
\[
  f\left(z+i{T\over 2a}\right)=f(z),~~~
  g\left(\bar{z}-i{T\over 2a}\right)=g(\bar{z});
\]

ii) existence of poles at $z,\bar{z}=ilT/2a$,  $l=0,\pm 1,\pm 2,\ldots$
(recall that in the dilute gas approximation, the periodic instanton is
a chain of instantons sitting at these points);

iii) regularity of the function $\chi$, eq.(\ref{gensol}), everywhere in
Euclidean space-time;

iv) reality,
\[
  g(\bar{z})=(f(z))^*,
\]
ensuring that $\chi$ is real in Euclidean space-time.

v) reality of $\chi$  also in Minkowski time (on both
Minkowskian parts of the contour of fig.1).

These conditions are sufficient to determine $f$ and
$g$. We find
\[
  f(z)={B\over \tanh(\mu z)},~~~g(\bar{z})={B\over \tanh(\mu \bar{z})},
\]
where
\[
\mu={2\pi a\over T}
\]
and $B$ is yet unknown parameter.
 The field configuration corresponding to this
choice is
\begin{equation}
  \phi=-{v\over\lambda}\ln\left[{1\over\mu}
      \left(B\cosh\mu z\cosh\mu\bar{z}+
      {1\over B}\sinh\mu z\sinh\mu\bar{z}\right)\right].
  \label{period}
\end{equation}
The values of $T$ (or $\mu$) and $B$ have to be found by matching this
solution to the massive free field and requiring that the  energy of
this solution takes a given value $E$. The contour plots of the configuration
(\ref{period}) for different values of $B$ is presented in fig.8.

First consider the case $B\ll 1$.
At small $z,\bar{z}$, one has
\[
  \phi=-{v\over\lambda}\ln\left({B\over\mu}+
        {\mu\over B}\bar{z}z\right),
\]
which coincides with eq.(\ref{Liouv_sol}) for $c=B/\mu$. So,
$B$ is the  diluteness  parameter  and
$B\ll 1$ corresponds to the case of the dilute gas.

 In what follows
we concentrate on the case $B\sim 1$, when the
improved dilute gas approximation
is not valid. Let us assume that the period $T$ is small enough,
$T\ll 1/m$ (this assumption will be justified a posteriori). Then the
field becomes free and massless at $T\ll x\ll 1/m$. Indeed, in this
region we have $\mu z\gg 1$ and the configuration (\ref{period}) is a
time-independent solution to the massless free field equation,
\[
  \phi=-{v\over\lambda} \ln\left[{1\over
       4\mu}\left(B+{1\over B}\right)\right]-
       {2v\mu\over\lambda}|x|.
\]
We have to match this configuration to the time-independent
  solution to the
massive free equation,
$\phi\propto\e^{-m|x|}$. This matching results in the following
relation,
\begin{equation}
  \mu=\e^{\lambda}(1+O(\lambda^{-1})),
  \label{mu}
\end{equation}
and the field outside the core is
\begin{equation}
\phi=v\e^{-m|x|}.
\label{massives}
\end{equation}

The above formulas explicitly define the periodic instanton at given
period $T$.
At $B=1$, the field (\ref{period}) is independent of
$\tau$, and describes the core of the sphaleron,
\[
  \phi=-{v\over\lambda}\ln\left[{1\over\mu}\cosh\mu(z+\bar{z})\right].
\]
The values of the parameter $B$ of order 1 correspond to
energies at which the dilute instanton gas approximation does not work, i.e.,
$1-E/E_{sph}\sim 1/\lambda$.
The period is approximately constant in this  energy interval,
\begin{equation}
  T={2\pi a\over\mu}={4\pi\over\lambda m}(1+O(\lambda^{-1})).
  \label{Tmu}
\end{equation}
Since $T$ does not change
much in this region, we make use of the relation
$\partial F/\partial E=T/g^2$, to obtain  that eq.(\ref{FEperiod}) is
correct, with the accuracy $1/\lambda^2$,
at all energies including  the region close to the sphaleron mass.

To find the value of $B$, we make use of
 the following trick. Let us
calculate explicitly the value of $F$ for the
configuration (\ref{period}).  We write
\[
  F=2S-ET=2\int\limits_0^{T/2}dt\int\limits_{-\infty}^{\infty}dx
  \left(\half(\partial_{\tau}\phi)^2+\half(\partial_x\phi)^2+V(\phi)\right)
\]
\[
  -2\int\limits_0^{T/2}dt\int\limits_{-\infty}^{\infty}dx
  \left(-\half(\partial_{\tau}\phi)^2+\half(\partial_x\phi)^2+V(\phi)\right)
\]
\[
  =2\int\limits_0^{T/2}dt\int\limits_{-\infty}^{\infty}dx
   (\partial_{\tau}\phi)^2.
\]
The result of the integration for the actual field,
 eq.(\ref{period}), is
\[
  F={8\pi v^2\over\lambda^2}\ln\left[\half\left(B+{1\over B}\right)
    \right].
\]
We now compare this result and eq.(\ref{FEperiod}) and find
\[
  1-{E\over E_{sph}}={2\over\lambda}\ln\left[\half\left(B+{1\over B}
  \right) \right]
\]
This relation determines the value of $B$ at given energy,
so that all parameters of the solution are now defined. Notice that
$B\sim1$ indeed corresponds to $(E_{sph}-E)/E_{sph}\sim 1/\lambda$.
Note also that eq.(\ref{Tmu}) justifies the
assumption
that $T\ll 1/m$, so our calculation is reliable at all energies up to
$E_{sph}$.

The above procedure for obtaining the relevant classical solution is
likely to work at $\theta\sim 1/\lambda$, i.e., for processes with the
number of particles of order $n_{sph}$. Unfortunately, this procedure
does not work in the most interesting case $\theta\sim 1$ (i.e., for
small number of particles in the initial state): the size of the
solution to the Liouville equation very rapidly increases with energy
and becomes of order $1/m$ essentially at the same energy as one given
by eq.(\ref{critE}). The calculations leading to the latter negative
conclusion are not illuminating and we do not present them here.

\section{Induced vacuum decay}

In this section we briefly discuss a process which is closely related to the
shadow processes, namely, the decay of the false vacuum induced by initial
particles. The most probable process of the induced
vacuum decay at a given energy and number of initial particles is described
by the solution to the same boundary
 value problem, eqs.(\ref{bvp1}) and (\ref{bvp2}),
but with the condition that the field
 in the final state (the part $CD$ of the contour of fig.1)
describes an expanding bubble with the large value of the scalar field inside.
 In the dilute gas approximation, the
configuration is a chain of instantons located at $(lT,0)$ with intensities
$\exp(-|l|\theta)$ (fig.9). The decay rate is again exponential,
\[
  \Gamma \sim \exp(-F),
\]
where $F$ is calculated according to eq.(\ref{FS}). As is seen from
fig.9, the induced vacuum decay can be considered as a half--instanton
process.

The solution to the boundary value problem can be obtained, in the
improved dilute instanton gas approximation, in a complete analogy to
Sect.3.
The details of the calculation are not very instructive, so we present
here our results for the rates only.

i) In the case $\theta\sim 1/\lambda$, the function $F(E,n)$ of the vacuum
decay
is exactly  half of the function $F$ for the shadow process at the same
energy and the same number of particles. The region of validity of the
improved dilute gas approximation is also the same.

ii) For parametrically larger $\theta$, $\theta\sim 1$, the formula analogous
to eq.(\ref{Fkappa}) reads
\[
  F = S_0\left[1-2{\e^{-\theta}\over 1-\e^{-\theta}}
      \left(1+{\theta\over 1-\e^{-\theta}}\right)
      {1\over\lambda}\ln{E\over E_{sph}}\right],
\]
where $\theta$ is again a function of particle number and energy,
\[
  \theta=\arccosh\left(1+{2\pi^2\over\lambda^2}{n_{sph}\over n}
         \ln{E\over E_{sph}}\right).
\]

In the limit $\theta\to\infty$, all  instantons, excluding the central
one, can be neglected, and the function $F$ is equal to its zero--energy
value, $S_0$. So, we find that at energies where the improved dilute
 gas approximation is valid, the vacuum decay is not enhanced by a small
(of order $v^2/\lambda$) number of initial particles. The region of
validity of the dilute gas again extends to a certain critical energy,
that depends on the number of initial particles and is of order
\[
E_{crit}\sim\e^{\lambda}E_{sph}
\]
 in the limit $\theta\to\infty$, i.e, when the number of initial
 particles is small.

\section{$\phi^N$ model}

In this section we consider another example of the (1+1) dimensional scalar
theory with the quadratic potential with a cliff. Namely, we take
the scalar
potential to have the following form,
\begin{equation}
  V(\phi) = {m^2\phi^2\over 2} - {m^2\over 2v^N}\phi^{N+2},
\label{Npot}
\end{equation}
where $N$ is a large number
playing the role of $\lambda$ in the potential model.
 The qualitative behavior of the potential is
the same as in the exponential model.

The sphaleron field outside the core is precisely the same as in the
exponential model. In particular, the sphaleron energy and the number
of particles are given by eqs.(\ref{Esph}) and (\ref{Nsph}),
respectively.

The technique developed for the exponential model to describe the
instanton and improved dilute instanton gas requires a
slight modification
in the power model.
Let us find out the instanton in this model. The Euclidean field
equation has the form
\begin{equation}
  \partial_{\mu}^2\phi = m^2\phi - (N+2){m^2\over 2v^N}\phi^{N+1}.
  \label{phiNeq}
\end{equation}
First let us consider the region
$r\ll m^{-1}$, where the mass term in the field equation can be neglected.
Though the nonlinear term in eq.(\ref{phiNeq}) does not have the exponential
form, in this region eq.(\ref{phiNeq}) reduces to the Liouville
equation. To see this, let us make the following change of variables
\begin{equation}
  \phi = \phi_0\left(1-{2\chi\over N}\right),
  \label{phichi}
\end{equation}
where $\phi_0=\phi(0)$ is a yet undetermined parameter,
 and consider only such $r$ for which $\chi(r)\ll N$
(i.e. $1-\phi(r)/\phi_0\ll 1$). Eq.(\ref{phiNeq}) obviously reduces to
the Liouville equation (recall that $N\gg 1$),
\begin{equation}
  \partial_{\mu}^2\chi = {4\over r_0^2}\e^{-2\chi},
  \label{phiNLiouv}
\end{equation}
where
\[
  r_0={4\over Nm}\left({\phi_0\over v}\right)^{-N/2}.
\]
The $O(2)$ symmetric solution to  eq.(\ref{phiNLiouv}) obeying the
condition $\chi(0)=0$ is
\begin{equation}
  \chi(r)=\ln\left(1+{r^2\over r_0^2}\right).
\label{Nchi}
\end{equation}
We now match the solution, eqs.(\ref{phichi}),(\ref{Nchi}), to free
massive field at $m^{-1}\gg r\gg r_0$,
\begin{equation}
  \phi(r)=\phi_0\left(1-{2\over N}\ln\left(1+{r^2\over r_0^2}\right)\right)
	 =\alpha K_0(mr),
  \label{phiNmatch}
\end{equation}
and obtain
\[
  \phi_0 = v\e^{1/2}, ~~~r_0\sim m^{-1}\e^{-N/4},
\]
\[
  \alpha = {4\e^{1/2}v\over N}.
\]
Notice that the size of the instanton core, $r_0$, is indeed
exponentially smaller than $m^{-1}$.
The instanton action is equal to
\[
  S_0 = {4\pi\e v^2\over N}.
\]

Given the instanton field, it is straightforward to evaluate the leading
order instanton contribution to the two-particle total cross section for
the shadow process. We find
\begin{equation}
  F(E)^{leading}= 2S_0\left(1-\frac{4}{N}\ln\frac{E}{E_{sph}}\right)
\label{PX}
\end{equation}
so that the energy scale relevant to the two-particle processes is again
exponentially large as compared to the sphaleron mass,
\[
  E_0\sim E_{sph}\e^{N/4}
\]

Now let us consider the improved dilute instanton gas.
The field outside the instanton core is still given by
eq.(\ref{dilgas}), while the field inside the core is determined by
eqs.(\ref{phichi}),(\ref{Nchi})
where $\phi_0=\phi(0)$ is yet unknown.
The analog of the matching condition, eq.(\ref{phiNmatch}),
is that at $m^{-1}\gg r\gg r_0$ the following relation should hold,
\[
  \phi(r)=\phi_0\left(1-{2\over N}\ln\left(1+{r^2\over r_0^2}\right)\right)
	 =\alpha(K_0(mr)+f),
\]
where $f$ is defined by the same formula as in the case of the exponential
model, eq.(\ref{f}). One obtains
\[
  \phi_0 = v\exp\left({1\over 2}-{2\over N}f\right),
  ~~~r_0\sim m^{-1}\exp\left(-{N\over 4}+f\right).
\]
\begin{equation}
  \alpha = {4v\over N}\exp\left({1\over 2}-{2\over N}f\right).
  \label{alphaphiN}
\end{equation}
It is worth noting that the parameter $\alpha$ depends on $f$, i.e., on
the parameters $T$ and $\theta$, in contrast
to the exponential model, where $\alpha$ is constant. So, not only
inside the core, but also outside the core the field is sensitive to
 the distance
between the instantons.

In the case $\theta\sim 1/N$,
i.e., at $n/n_{sph}\sim 1$,
 one obtains from eqs.(\ref{fbeta}), (\ref{dil_E}),
(\ref{dil_n}) and (\ref{alphaphiN})  the following expressions
for the energy and the number of initial particles,
\[
  E = x^2\exp(1-xf_1(\beta))f_2(\beta)E_{sph},
\]
\[
  n = x^2\exp(1-xf_1(\beta))f_3(\beta)n_{sph},
\]
where
\[
  x=\frac{8\pi}{NmT}
\]
\[
 \beta= \frac{\theta}{mT}
\]
and $f_1$, $f_2$ and $f_3$ are the same functions as defined
in Sect.3.
 The calculations analogous to those of the
exponential model give for the action
\[
  S = {4\pi\e v^2\over N}\exp(1-xf_1(\beta)),
\]
so  the function $F=2S-ET-n\theta$  has the following form,
\[
  F = {8\pi\e v^2\over N}(1-xf_1(\beta))\exp(1-xf_1(\beta)).
\]
Though these formulas
are slightly more complicated than eqs.(\ref{Enbeta}) and (\ref{Fbeta}), the
qualitative behavior of the function $F(E,n)$ is the same as in the case
of the exponential model. In other words, the behavior shown in fig.5
is qualitatively correct in the power model as well. The improved
instanton gas approximation is valid until the probability becomes
parametrically less suppressed as compared to the instanton,
\[
F\sim \frac{S_0}{N},
\]
in a complete analogy to the exponential model.

 In the regime $\theta\sim 1$, i.e., at $n/n_{sph}\sim 1/N$, the
 interesting energies are exponentially large, so that
\[
  \frac{1}{N}\ln\frac{E}{E_{sph}}=\kappa\sim 1.
\]
The formula analogous to eq.(\ref{Fkappa}) is
\[
  F=2S_0\exp\left(-4{1+3\e^{-\theta}\over 1-\e^{-\theta}}\kappa\right)
    \left[1-\frac{16\theta\e^{-\theta}}{ (1-\e^{-\theta})^2}\kappa\right],
\]
where $\theta$ is to be determined
 from a relation analogous to eq.(\ref{nkappa}),
\[
  n={64\pi^2\e n_{sph}\over N}{e^{-\theta}\over (1-\e^{-\theta})^2}\kappa
    \exp\left(-4{1+3\e^{-\theta}\over 1-\e^{-\theta}}\kappa\right).
\]
The size of the core is
\[
  r_0 \sim {1\over m}\exp\left(-{N\over 4}+N{1+3\e^{-\theta}\over
           1-\e^{-\theta}}\kappa\right).
\]
In the limit $\theta\to\infty$, i.e., $n/n_{sph}\to 0$,
 the behavior of the function $F$ is as follows,
\begin{equation}
  F = 2S_0 \exp\left(-{4\over N}\ln{E\over E_{sph}}\right) =
      2S_0 \left({E\over E_{sph}}\right)^{-4/N}.
  \label{FphiN}
\end{equation}
Note that in the $\phi^N$ model, the function $F(E/E_{sph})$
contains higher order
terms in $N^{-1}\ln(E/E_{sph})$, in contrast to the exponential model,
where all higher terms vanish and the result coincides with the
one--instanton expression.

Let us explain the latter result.
 Recall that in the exponential model only the
field inside the instanton core is modified by the presence of other
instantons, while the linear tail of an instanton in the chain is the
same as that
of a single instanton (the parameter $\alpha$ is constant
independent of $T$ and $\theta$). If the cores do
not overlap, the field configuration outside the cores is the same as in the
ordinary dilute instanton gas approximation. The dependence of $E$ and $n$
on $T$ and $\theta$, eqs.(\ref{dil_E}) and (\ref{dil_n}), thus coincides
with the formulas of the ordinary dilute instanton gas approximation. Since
the function $F$ can be recovered from eq.(\ref{derivF}), it is not
surprising that the improved dilute instanton gas implies the same result
for $F$ as the lowest order of the perturbation theory around a single
instanton. In the $\phi^N$ model, on the contrary, the intensity of the linear
tail depends on $T$ and $\theta$ (see eq.(\ref{alphaphiN})), so the
 correction to the one--instanton formula for the function $F$ is large.

In the limit $\theta\to\infty$, the improved dilute gas
approximation breaks downs at $E_{crit}\sim e^{N/8}E_{sph}$, where
$F=2S_0/\sqrt{\e}$. As in the exponential model, the function $F$ is reduced
only by a factor of order one
 ( 2/3 in the exponential model and $\e^{-1/2}$ in the power model).

\section{Conclusions}

In this paper we have found the solution to the classical boundary value
problem for certain (1+1) dimensional field theories with the potential
having the form of a quadratic potential with a cliff. The presence of an
additional large parameter in the model allowed us to deal with energies
comparable to or much larger than the sphaleron mass. The solution, which
describes the shadow processes, has been found reliably within the improved
dilute
instanton gas approximation at large energies where the ordinary dilute
instanton gas approximation does not work. We have found that for
the number of initial particles, $n$, of order
$n_{sph}$, the function $F(E/E_{sph},n)$ monotonously decreases from $2S_0$ at
$E=0$ and becomes of order $S_0/\lambda$
(or $S_0/N$) at some $E_{crit}$, where the
improved dilute gas approximation fails. For $n$ of order
$n_{sph}/\lambda$ (or $n_{sph}/N$ in the $\phi^N$ model), including the limit
$n\to 0$ which is smooth, the improved dilute gas approximation breaks down
at energies where the function $F$ is still of order $S_0$, and the
instanton--like transitions are still suppressed. So, although in our models we
can go  beyond the ordinary perturbative theory around the instanton,
 the possibility of large probability of the
instanton--like transitions induced by two energetic particles remains
an open question even in these models.

The authors are indebted to A.N.Kuznetsov and P.G.Tinyakov for numerous helpful
discussions at all stages of this work. We are indebted to
T.Banks and V.Petrov for the discussion of the results. V.R. thanks
Aspen Center for Physics and Rutgers University for hospitality.
 The work of D.T.S. is supported in part by
  the Russian Foundation for Fundamental Research (project 93-02-3812)
and  by the Weingart
Foundation through a cooperative agreement with the Department of Physics at
UCLA.

\newpage
\begin{picture}(300,200)
\put (50,100){\vector(1,0){200}}
\put (150,40){\vector(0,1){130}}
\thicklines
\put (70,140){\line(1,0){80}}
\put (150,140){\line(0,-1){40}}
\put (150,100){\line(1,0){80}}
\put (70,145){$A$}
\put (135,145){$B$}
\put (135,105){$C$}
\put (220,105){$D$}
\put (157,135){$T/2$}
\put (155,165){Re~$t$}
\put (240,87){Im~$t$}
\put (140,0){Fig.1}
\end{picture}

\newpage
\begin{picture}(300,400)
\put (50,150){\vector(1,0){200}}
\put (150,40){\vector(0,1){300}}
\thicklines
\put (150,200){\circle*{10}}
\put (150,100){\circle*{10}}
\put (150,230){\circle*{7}}
\put (150,70){\circle*{7}}
\put (150,280){\circle*{7}}
\put (150,310){\circle*{5}}
\put (140,230){\makebox[1pt][r]{e$^{-\theta}I$}}
\put (140,100){\makebox[1pt][r]{$I$}}
\put (140,194){\makebox[1pt][r]{$I$}}
\put (140,280){\makebox[1pt][r]{e$^{-\theta}I$}}
\put (140,310){\makebox[1pt][r]{e$^{-2\theta}I$}}
\put (140,70){\makebox[1pt][r]{e$^{-\theta}I$}}
\put (160,194){\makebox[1pt][l]{$T_1$}}
\put (160,213){\makebox[1pt][l]{$T/2$}}
\put (160,230){\makebox[1pt][l]{$T-T_1$}}
\put (160,280){\makebox[1pt][l]{$T+T_1$}}
\put (160,310){\makebox[1pt][l]{$2T-T_1$}}
\put (160,100){\makebox[1pt][l]{$-T_1$}}
\put (160,70){\makebox[1pt][l]{$-T+T_1$}}
\put (70,215){\line(1,0){80}}
\put (150,215){\line(0,-1){65}}
\put (150,150){\vector(1,0){100}}
\put (70,220){$A$}
\put (140,155){$C$}
\put (240,155){$D$}
\put (140,0){Fig.3}
\end{picture}

\newpage
\begin{picture}(200,200)
\put(100,130){\vector(1,0){200}}
\put(120,50){\vector(0,1){100}}
\put(260,130){\line(0,-1){5}}
\put(120,60){\line(1,0){5}}
\bezier{500}(120,60)(190,130)(260,130)
\put(124,150){$-F(E)$}
\put(300,136){$E$}
\put(255,136){$E_{sph}$}
\put(90,56){$-2S_0$}
\put(109,136){0}
\put(200,0){Fig.4}
\end{picture}

\newpage
\begin{picture}(400,300)
\put(30,200){\circle{20}}
\put(20,200){\line(1,1){10}}
\put(21,196){\line(1,1){13}}
\put(23,193){\line(1,1){14}}
\put(26,191){\line(1,1){13}}
\put(30,190){\line(1,1){10}}
\put(2,228){\vector(1,-1){11}}
\put(2,172){\vector(1,1){11}}
\put(13,217){\line(1,-1){10}}
\put(13,183){\line(1,1){10}}
\put(39,204){\vector(2,1){27}}
\put(39,196){\vector(2,-1){27}}
\put(36,208){\vector(3,4){18}}
\put(36,192){\vector(3,-4){18}}
\put(40,200){\vector(1,0){30}}

\put(85,196){=}

\put(130,200){\circle{20}}
\put(128,196){I}
\put(102,228){\vector(1,-1){11}}
\put(102,172){\vector(1,1){11}}
\put(113,217){\line(1,-1){10}}
\put(113,183){\line(1,1){10}}
\put(139,204){\vector(2,1){27}}
\put(139,196){\vector(2,-1){27}}
\put(136,208){\vector(3,4){18}}
\put(136,192){\vector(3,-4){18}}
\put(140,200){\vector(1,0){30}}

\put(185,196){+}

\put(230,200){\circle{20}}
\put(228,196){I}
\put(202,228){\vector(1,-1){11}}
\put(202,172){\vector(1,1){11}}
\put(213,217){\line(1,-1){10}}
\put(213,183){\line(1,1){10}}
\put(280,200){\circle{20}}
\put(278,196){I}
\put(240,200){\line(1,0){30}}
\bezier{100}(239,204)(255,213)(271,204)
\bezier{100}(239,196)(255,187)(271,196)
\bezier{100}(236,208)(255,233)(274,208)
\bezier{100}(236,192)(255,167)(274,192)
\put(289,204){\vector(2,1){27}}
\put(289,196){\vector(2,-1){27}}
\put(286,208){\vector(3,4){18}}
\put(286,192){\vector(3,-4){18}}
\put(290,200){\vector(1,0){30}}

\put(85,116){+}

\put(130,120){\circle{20}}
\put(128,116){I}
\put(102,148){\vector(1,-1){11}}
\put(102,92){\vector(1,1){11}}
\put(113,137){\line(1,-1){10}}
\put(113,103){\line(1,1){10}}
\put(140,120){\line(1,0){30}}
\bezier{100}(139,124)(155,133)(171,124)
\bezier{100}(139,116)(155,107)(171,116)
\bezier{100}(136,128)(155,153)(174,128)
\bezier{100}(136,112)(155,87)(174,112)
\put(180,120){\circle{20}}
\put(178,116){I}
\put(190,120){\line(1,0){30}}
\bezier{100}(189,124)(205,133)(221,124)
\bezier{100}(189,116)(205,107)(221,116)
\bezier{100}(186,128)(205,153)(224,128)
\bezier{100}(186,112)(205,87)(224,112)
\put(230,120){\circle{20}}
\put(228,116){I}
\put(239,124){\vector(2,1){27}}
\put(239,116){\vector(2,-1){27}}
\put(236,128){\vector(3,4){18}}
\put(236,112){\vector(3,-4){18}}
\put(240,120){\vector(1,0){30}}
\put(285,116){+}
\put(300,116){$\cdots$}
\put(155,20){Fig.6}
\end{picture}

\newpage
\begin{picture}(300,400)
\put (50,150){\vector(1,0){200}}
\put (150,40){\vector(0,1){300}}
\thicklines
\put (150,170){\circle*{10}}
\put (150,130){\circle*{10}}
\put (150,200){\circle*{10}}
\put (150,100){\circle*{10}}
\put (150,220){\circle*{7}}
\put (150,250){\circle*{7}}
\put (150,290){\circle*{7}}
\put (140,94){\makebox[1pt][r]{$I$}}
\put (140,124){\makebox[1pt][r]{$I$}}
\put (140,164){\makebox[1pt][r]{$I$}}
\put (140,194){\makebox[1pt][r]{$I$}}
\put (140,220){\makebox[1pt][r]{e$^{-\theta}I$}}
\put (140,250){\makebox[1pt][r]{e$^{-\theta}I$}}
\put (140,290){\makebox[1pt][r]{e$^{-\theta}I$}}
\put (160,164){\makebox[1pt][l]{$T_1$}}
\put (160,124){\makebox[1pt][l]{$-T_1$}}
\put (160,194){\makebox[1pt][l]{$T_2$}}
\put (160,94){\makebox[1pt][l]{$-T_2$}}
\put (160,250){\makebox[1pt][l]{$T-T_1$}}
\put (160,290){\makebox[1pt][l]{$T+T_1$}}
\put (160,220){\makebox[1pt][l]{$T-T_2$}}
\put (70,210){\line(1,0){80}}
\put (150,210){\line(0,-1){60}}
\put (150,150){\vector(1,0){100}}
\put (140,0){Fig.7}
\end{picture}

\newpage
\begin{picture}(300,300)(0,-20)
\put (50,150){\vector(1,0){200}}
\put (150,40){\vector(0,1){260}}
\thicklines
\put (150,150){\circle*{10}}
\put (150,210){\circle*{7}}
\put (150,90){\circle*{7}}
\put (150,270){\circle*{5}}
\put (140,160){\makebox[1pt][r]{$I$}}
\put (140,210){\makebox[1pt][r]{e$^{-\theta}I$}}
\put (140,270){\makebox[1pt][r]{e$^{-2\theta}I$}}
\put (140,90){\makebox[1pt][r]{e$^{-\theta}I$}}
\put (160,210){\makebox[1pt][l]{$T$}}
\put (160,178){\makebox[1pt][l]{$T/2$}}
\put (160,270){\makebox[1pt][l]{$2T$}}
\put (160,90){\makebox[1pt][l]{$-T$}}

\put (70,180){\line(1,0){80}}
\put (150,180){\line(0,-1){30}}
\put (150,150){\vector(1,0){100}}
\put (140,0){Fig.9}
\end{picture}

\newpage
Figure captions:

1. The contour in the complex time plane, on which
 the boundary value problem is formulated.

2. The potential $V(\phi)$.

3. The schematic plot of the
solution to the boundary value problem in the improved dilute instanton
gas approximation. $I$ stand for the instanton.

4. The function $F(E)$ for the periodic instanton.

5. The function $F(E,n)$ for $\nu=n/n_{sph}\sim 1$. In the regions where
$F\sim S_0/\lambda$, the curves are unreliable.

6. Multi--instanton contributions to $2\to${\em many} amplitudes.

7. The two--instanton solution to the boundary value problem.

8. The contour plots of the periodic instanton solution, eq.(\ref{period}),
for differnent values of $B$. Larger $B$ correspond to higher energies.
As $B$ increases from 0 ($E=0$) to 1 ($E=E_{sph}$), the configuration
evolves from a chain of instantons to the sphaleron.

9. The configuration decribing the induced vacuum decay
in the improved dilute instanton gas approximation.


\begin{thebibliography}{99}

\bibitem{Ringwald}
     A.Ringwald, {\em Nucl.Phys.} {\bf B330} (1990) 1;\\
     O.Espinosa, {\em Nucl.Phys.} {\bf B334} (1990) 310.
\bibitem{MVV} L.McLerran, A.I.Vainshtein and M.B.Voloshin
     {\em Phys.Rev.} {\bf D42} (1990) 171.
\bibitem{Mrev}
    M.Mattis, {\em Phys.Rep.} {\bf 214} (1992) 159.
\bibitem{Trev}
    P.G.Tinyakov, {\em Int.J.Mod.Phys.} {\bf A8} (1993) 1823.
\bibitem{Zakharov}
    V.I.Zakharov,  {\em Classical Corrections to Instanton-Induced
    Amplitudes}, preprint TPI-MINN-90/7-T, 1990;
    {\em Nucl.Phys.} {\bf B371} (1992) 637.
\bibitem{KRT}
       S.Yu.Khlebnikov, V.A.Rubakov and P.G.Tinyakov, {\em
     Mod.Phys.Lett.} {\bf A5} (1990) 1983; {\em Nucl.Phys.}
     {\bf B350} (1991) 441.
\bibitem{Porrati}
       M.Porrati, {\em Nucl.Phys.} {\bf B347} (1991) 371.
\bibitem{KhozeRingwald}
    V.V.Khoze and A.Ringwald, {\em Nucl.Phys.} {\bf B355} (1991) 351.
\bibitem{MuellerDiakonovPetrov}
    A.H.Mueller, {\em Nucl.Phys.} {\bf B364} (1991) 109.\\
    D.I.Diakonov and V.Yu.Petrov, {\em Baryon Number Non--Conservation in
    Processes at High Energies}, In: Proceedings of XXVI Winter School, LINP,
    Leningrad, 1991.\\
    P.B.Arnold and M.P.Mattis, {\em Mod.Phys.Lett.} {\bf A6} (1991) 2095.
\bibitem{Mueller2}
    A.H.Mueller, {\em Nucl.Phys.} {\bf B348} (1991) 310;
    {\bf B353} (1991) 44.
\bibitem{RT}
  V.A.Rubakov and P.G.Tinyakov, {\em Phys.Lett.} {\bf B279} (1992) 165.
\bibitem{Tinyakov}
  P.G.Tinyakov, {\em Phys.Lett.} {\bf B284} (1992) 410.
\bibitem{Khlebnikov}
   S.Khlebnikov, {\em Variational approach to multiparticle production},
   preprint UCLA/91/TEP/38(1991).
\bibitem{DPN}
   D.Diakonov and V. Petrov, {\em Non-perturbative isotropic
   multi-particle production in Yang-Mills theory}, preprint RUB - TPII
   - 52/93(1993)
\bibitem{VoloshinN}
   M.B.Voloshin, {\em Catalyzed decay of false vacuum in four
   dimensions}, preprint TPI - MINN - 93/37 - T(1993)
\bibitem{RST}
  V.A.Rubakov, D.T.Son and P.G.Tinyakov,
  {\em Phys.Lett.} {\bf 287B} (1992) 342.
\bibitem{Mueller3}
    A.H.Mueller, {\em Nucl.Phys.} {\bf B401} (1993) 93.
\bibitem{Coleman}
    S.Coleman, {\em Phys.Rev.} {\bf D15} (1977) 2929;\\
    C.Callan and S.Coleman {\em Phys.Rev.} {\bf D16} (1977) 1762.
\bibitem{Hsu}
    S.D.H.Hsu, {\em Phys.Lett.} {\bf B261} (1991) 81.
\bibitem{Voloshin}
    M.B.Voloshin, {\em Nucl.Phys.} {\bf B363} (1991) 425.
\bibitem{Zakharov1}
     V.Zakharov, {\em Nucl.Phys} {\bf B353} (1991) 683.
\bibitem{M&S}
     M.Maggiore and M.Shifman, {\em Nucl.Phys.} {\bf B365} (1991) 161;
     {\bf B371} (1991) 177.
\bibitem{KRTperiod}
     S.Yu.Khlebnikov, V.A.Rubakov and P.G.Tinyakov, {\em Nucl.Phys.}
 {\bf B367} (1991) 334.

\end{thebibliography}
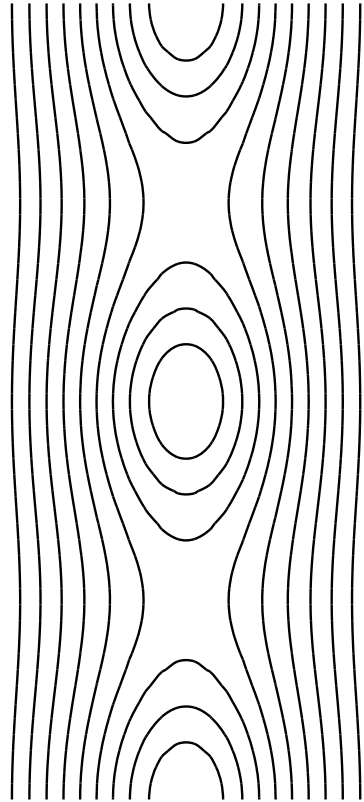
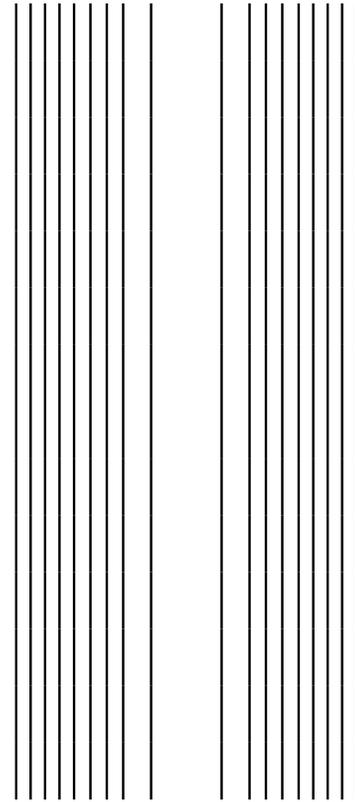
\end{document}